\RequirePackage{lineno_babar}
\documentclass[11pt]{article}
\usepackage{graphicx}

\newcommand{\BABARPubYear}    {07}

\newcommand{\BABARConfNumber} {006}
\newcommand{\SLACPubNumber}{12696}

\long\def\inst#1{\par\nobreak\kern 4pt\nobreak
    {\it #1}\par\vskip 10pt plus 3pt minus 3pt}

\usepackage{graphicx}
\usepackage{dcolumn}
\usepackage{amsmath}
\usepackage{epsfig}

\RequirePackage{xspace}

\usepackage{relsize}
\def\babar{\mbox{\slshape B\kern-0.1em{\smaller A}\kern-0.1em
    B\kern-0.1em{\smaller A\kern-0.2em R}}}

\def\epem       {\ensuremath{e^+e^-}\xspace}

\def\qqbar {\ensuremath{q\overline q}\xspace}

\def\piz   {\ensuremath{\pi^0}\xspace}

\def\pip   {\ensuremath{\pi^+}\xspace}
\def\pim   {\ensuremath{\pi^-}\xspace}
\def\pipi  {\ensuremath{\pi^+\pi^-}\xspace}

\def\Kbar  {\kern 0.2em\overline{\kern -0.2em K}{}\xspace}

\def\Kz    {\ensuremath{K^0}\xspace}
\def\Kzb   {\ensuremath{\Kbar^0}\xspace}
\def\KzKzb {\ensuremath{\Kz \kern -0.16em \Kzb}\xspace}
\def\Kp    {\ensuremath{K^+}\xspace}
\def\Km    {\ensuremath{K^-}\xspace}

\def\KpKm  {\ensuremath{\Kp \kern -0.16em \Km}\xspace}
\def\KS    {\ensuremath{K^0_{\scriptscriptstyle S}}\xspace}

\def\Dbar    {\kern 0.2em\overline{\kern -0.2em D}{}\xspace}

\def\Dz      {\ensuremath{D^0}\xspace}
\def\Dzb     {\ensuremath{\Dbar^0}\xspace}
\def\DzDzb   {\ensuremath{\Dz {\kern -0.16em \Dzb}}\xspace}
\def\Dp      {\ensuremath{D^+}\xspace}
\def\Dm      {\ensuremath{D^-}\xspace}

\def\DpDm    {\ensuremath{\Dp {\kern -0.16em \Dm}}\xspace}

\def\Bbar    {\kern 0.18em\overline{\kern -0.18em B}{}\xspace}

\def\BB      {\ensuremath{B\Bbar}\xspace} 
\def\Bz      {\ensuremath{B^0}\xspace}
\def\Bzb     {\ensuremath{\Bbar^0}\xspace}
\def\BzBzb   {\ensuremath{\Bz {\kern -0.16em \Bzb}}\xspace}
\def\Bu      {\ensuremath{B^+}\xspace}
\def\Bub     {\ensuremath{B^-}\xspace}
\def\Bp      {\ensuremath{\Bu}\xspace}
\def\Bm      {\ensuremath{\Bub}\xspace}

\def\BpBm    {\ensuremath{\Bu {\kern -0.16em \Bub}}\xspace}

\mathchardef\Upsilon="7107
\def\Y#1S{\ensuremath{\Upsilon{(#1S)}}\xspace}

\def\FourS {\Y4S}

\mathchardef\Deltares="7101
\mathchardef\Xi="7104
\mathchardef\Lambda="7103
\mathchardef\Sigma="7106
\mathchardef\Omega="710A

\def\Deltabar{\kern 0.25em\overline{\kern -0.25em \Deltares}{}\xspace}
\def\Lbar{\kern 0.2em\overline{\kern -0.2em\Lambda\kern 0.05em}\kern-0.05em{}\xspace}
\def\Sigbar{\kern 0.2em\overline{\kern -0.2em \Sigma}{}\xspace}
\def\Xibar{\kern 0.2em\overline{\kern -0.2em \Xi}{}\xspace}
\def\Obar{\kern 0.2em\overline{\kern -0.2em \Omega}{}\xspace}
\def\Nbar{\kern 0.2em\overline{\kern -0.2em N}{}\xspace}
\def\Xb{\kern 0.2em\overline{\kern -0.2em X}{}\xspace}

\def\BR         {{\ensuremath{\cal B}\xspace}}

\def\mes        {\mbox{$m_{\rm ES}$}\xspace}

\def\DeltaE     {\mbox{$\Delta E$}\xspace}

\newcommand{\tev}{\ensuremath{\mathrm{\,Te\kern -0.1em V}}\xspace}
\newcommand{\gev}{\ensuremath{\mathrm{\,Ge\kern -0.1em V}}\xspace}
\newcommand{\mev}{\ensuremath{\mathrm{\,Me\kern -0.1em V}}\xspace}
\newcommand{\kev}{\ensuremath{\mathrm{\,ke\kern -0.1em V}}\xspace}
\newcommand{\ev}{\ensuremath{\mathrm{\,e\kern -0.1em V}}\xspace}
\newcommand{\gevc}{\ensuremath{{\mathrm{\,Ge\kern -0.1em V\!/}c}}\xspace}
\newcommand{\mevc}{\ensuremath{{\mathrm{\,Me\kern -0.1em V\!/}c}}\xspace}
\newcommand{\gevcc}{\ensuremath{{\mathrm{\,Ge\kern -0.1em V\!/}c^2}}\xspace}
\newcommand{\mevcc}{\ensuremath{{\mathrm{\,Me\kern -0.1em V\!/}c^2}}\xspace}

\def\invfb   {\ensuremath{\mbox{\,fb}^{-1}}\xspace}

\def\mus  {\ensuremath{\rm \,\mus}\xspace}

\def\mus        {\ensuremath{\,\mu{\rm s}}\xspace}    

\def\ra                 {\ensuremath{\rightarrow}\xspace}
\def\to                 {\ensuremath{\rightarrow}\xspace}

\newcommand{\stat}{\ensuremath{\mathrm{(stat)}}\xspace}
\newcommand{\syst}{\ensuremath{\mathrm{(syst)}}\xspace}

\def\pep2{PEP-II}

\def\gsim{{~\raise.15em\hbox{$>$}\kern-.85em
          \lower.35em\hbox{$\sim$}~}\xspace}
\def\lsim{{~\raise.15em\hbox{$<$}\kern-.85em
          \lower.35em\hbox{$\sim$}~}\xspace}

\def\CP                {\ensuremath{C\!P}\xspace}

\xspace

\newcommand{\jprlBase}       {Phys.\ Rev.\ Lett.\xspace}
\newcommand{\jprBase}        {Phys.\ Rev.\xspace}
\newcommand{\jplBase}        {Phys.\ Lett.\xspace}

\newcommand{\nimBaseC}       {Nucl.\ Instr.\ and Methods\xspace}

\newcommand{\nim}       [1]  {\nimBaseC~{\bf #1}}

\newcommand{\jpl}       [1]  {\jplBase\ {\bf #1}}

\newcommand{\jprl}      [1]  {\jprlBase\ {\bf #1}}
\newcommand{\pr}        [1]  {\jprBase\ {\bf #1}}

\newcommand{\jprd}      [1]  {\jprBase\ D~{\bf #1}}

\def\jetset74   {\mbox{\tt Jetset \hspace{-0.5em}7.\hspace{-0.2em}4}\xspace}

\def\figurebox#1#2#3{%
    \def\arg{#3}%
    \ifx\arg\empty
    {\hfill\vbox{\hsize#2\hrule\hbox to #2{\vrule\hfill\vbox to #1{\hsize#2\vfill}\vrule}\hrule}\hfill}%
    \else
    {\hfill\epsfbox{#3}\hfill}%
    \fi}

\newcommand{\btodk}{$B^{-}\ra \Do K^{-}$}

\newcommand{\btodp}{$B^{-}\ra \Do \pi^{-}$}
\newcommand{\btodh}{$B^{-}\ra \Do h^{-}$}

\newcommand{\btdk}{\mbox{$B\ra \Do K$}}
\newcommand{\btdp}{\mbox{$B\ra \Do \pi$}}

\newcommand{\Do}{D^0}

\newcommand{\dzcppm}{\ensuremath{\Dz_{\CP\pm}}\xspace}

\newcommand{\bmtodzk}     {\ensuremath{\Bm \to \Dz \Km}\xspace}
\newcommand{\bmtodzbk}    {\ensuremath{\Bm \to \Dzb \Km}\xspace}
\newcommand{\bmtodcpk}    {\ensuremath{\Bm \to \Dz_{\CP\pm} \Km}\xspace}

\newcommand{\bmtodzp}     {\ensuremath{\Bm \to \Dz \pim}\xspace}

\newcommand{\bmtodcpp}	  {\ensuremath{\Bm \to \Dz_{\CP\pm} \pim}\xspace}

\newcommand{\bptodzbk}	  {\ensuremath{\Bp \to \Dzb \Kp}\xspace}
\newcommand{\bptodcpk}	  {\ensuremath{\Bp \to \Dz_{\CP\pm} \Kp}\xspace}

\newcommand{\bptodzbp}	  {\ensuremath{\Bp \to \Dzb \pip}\xspace}
\newcommand{\bptodcpp}	  {\ensuremath{\Bp \to \Dz_{\CP\pm}\pip}\xspace}

\newcommand{\btodcppmk}	  {\ensuremath{B \to \Dz_{\CP\pm} K}\xspace}

\newcommand{\btodcppmp}	  {\ensuremath{B \to \Dz_{\CP\pm} \pi}\xspace}

\def\kk{\ensuremath{\Kp \kern -0.16em \Km}\xspace}
\def\kspi0{\ensuremath{\KS \piz}\xspace}

\def\ksrho0{\ensuremath{\KS \rhoz}\xspace}

\def\btdp{\ensuremath{B^-\rightarrow D^0\pi^-}\xspace}
\def\btdk{\ensuremath{B^-\rightarrow D^0K^-}\xspace}

\begin{document}

{\pagestyle{empty}
\begin{flushright}
\babar-CONF-\BABARPubYear/\BABARConfNumber \\
SLAC-PUB-\SLACPubNumber \\
\end{flushright}

\par\vskip 1.5cm

\begin{center}
\large \bf
Measurement of the branching fraction ratios and {\boldmath \CP} asymmetries in {\boldmath $B^- \ra  D^{0}_{CP} K^-$} decays
\end{center}
\bigskip

\begin{center}
\large The \babar\ Collaboration\\
\mbox{ }\\
\today
\end{center}
\bigskip \bigskip

\begin{center}
\large \bf Abstract
\end{center}

We present a preliminary study of $B^-\to\Dz_{(\CP)}\pi^-$ and $B^-\to\Dz_{(\CP)}K^-$
decays, with the $\Dz_{(\CP)}$ reconstructed in the \CP-odd eigenstates 
$K_{s}\pi^{0}$, 
$K_{s}\omega$, in the \CP-even eigenstates \KpKm, \pipi, and in
 the (non-\CP) flavor eigenstate $K^\mp \pi^\pm$. 
Using a sample of about 382 million \FourS\ decays into $B\overline{B}$ pairs, collected
with the \babar\ detector operating at the \pep2 asymmetric-energy $B$ Factory at SLAC, we measure 
the ratios of the branching fractions
\begin{linenomath}
\begin{displaymath}
R_{\CP_{\pm}}\equiv \frac{\BR(B^-\to D^{0}_{\CP\pm}K^-)+ \BR(B^+\to D^{0}_{\CP\pm}K^+)}{[\BR(B^-\to D^{0}K^-)+\BR(B^+\to D^{0}K^+)]/2 } 
\end{displaymath}
\end{linenomath}
and the direct \CP asymmetry
\begin{linenomath}
\begin{displaymath}
A_{\CP_{\pm}}\equiv\frac{\BR(B^-{\ra}D^{0}_{\CP\pm}K^-)-\BR(B^+{\ra}D^{0}_{\CP\pm}K^+)}{\BR(B^-{\ra}D^{0}_{\CP\pm}K^-)+\BR(B^+{\ra}D^{0}_{\CP\pm}K^+)}.
\end{displaymath}
\end{linenomath}
The results are:
\begin{linenomath}
\begin{eqnarray}
      R_{\CP-} & = & \ \ 0.81 \pm 0.10 \stat \pm 0.05 \syst \nonumber \\
      R_{\CP+} & = & \ \ 1.07 \pm 0.10 \stat \pm 0.04 \syst \nonumber \\
      A_{\CP-} & = & -0.19 \pm 0.12 \stat \pm 0.02 \syst \nonumber \\
      A_{\CP+} & = & \ \ 0.35 \pm 0.09 \stat \pm 0.05 \syst \nonumber
\end{eqnarray}
\end{linenomath}
\vfill

\vspace{0.5cm}
\begin{center}
Contributed to the 
XXIII$^{\rm rd}$ International Symposium on Lepton and Photon Interactions at High~Energies, 8/13 -- 8/18/2007, Daegu, Korea
\end{center}

\vspace{0.5cm}
\begin{center}
{\em Stanford Linear Accelerator Center, Stanford University, 
Stanford, CA 94309} \\ \vspace{0.1cm}\hrule\vspace{0.1cm}
Work supported in part by Department of Energy contract DE-AC03-76SF00515.
\end{center}

\newpage
} 

\begin{center}
\small

The \babar\ Collaboration,
\bigskip

%
{B.~Aubert,}
{M.~Bona,}
{D.~Boutigny,}
{Y.~Karyotakis,}
{J.~P.~Lees,}
{V.~Poireau,}
{X.~Prudent,}
{V.~Tisserand,}
{A.~Zghiche}
\inst{Laboratoire de Physique des Particules, IN2P3/CNRS et Universit\'e de Savoie, F-74941 Annecy-Le-Vieux, France }
{J.~Garra~Tico,}
{E.~Grauges}
\inst{Universitat de Barcelona, Facultat de Fisica, Departament ECM, E-08028 Barcelona, Spain }
% ...

\end{center}\newpage

\section{INTRODUCTION}
  \label{sec:Introduction}

  A measurement of the processes \bmtodzk~\cite{conj}
  and \btodcppmk, where \dzcppm indicates the \CP-even or
  \CP-odd states $1/\sqrt{2}(D^0\pm \bar{D}^0)$,
  has been attracting
  the attention of theorists for the last fifteen years~\cite{gronau1991}. 
  These decay rates are fundamental
  ingredients in some of the proposed methods to extract the $\gamma$
  angle of the CKM matrix in a theoretically clean
  way. To this end, one needs to measure
  the two direct \CP\ asymmetries 
\begin{linenomath}
\begin{equation}
  A_{\CP\pm} \equiv
  \frac{\BR(\bmtodcpk)-\BR(\bptodcpk)}{\BR(\bmtodcpk)+\BR(\bptodcpk)}\ ,
  \label{eq:def_Acp}
\end{equation}
\end{linenomath}
  and the two ratios of charge-averaged branching fractions in \Dz decays to
  \CP eigenstates
\begin{linenomath}
\begin{equation}
  R_{\CP\pm} \equiv 
  \frac{\BR(\bmtodcpk)+\BR(\bptodcpk)}{\left[\BR(\bmtodzk)+\BR(\bptodzbk)\right]/2}.
  \label{eq:def_Rcp} 
\end{equation}
\end{linenomath}
  In fact, $\gamma$ is constrained by the following set of
  equations
  in the three unknowns $\gamma$, $\delta$, $r$: 
\begin{linenomath}
\begin{equation}
  \label{eq:rcp_vs_gamma}
  R_{\CP\pm}=1+r^2 \pm 2r\cos\delta \cos\gamma
\end{equation}
\end{linenomath}
\begin{linenomath}
\begin{equation}
  \label{eq:acp_vs_gamma}
  A_{\CP\pm}=\frac{\pm 2 r \sin\delta \sin\gamma}{1+r^2 \pm
  2r\cos\delta \cos\gamma},
\end{equation}
\end{linenomath}
  where $r \equiv
  \left|A(\bmtodzbk)/A(\bmtodzk)\right|\approx 
  \mathcal{O}(0.1)$ is the magnitude of the ratio of the amplitudes
  for \bmtodzbk and \bmtodzk
  and $\delta$ the difference between their strong phases.
  The asymmetries $A_{\CP\pm}$, in addition to being ingredients for the extraction 
  of $\gamma$, are of special relevance because they would indicate, if significantly
  different from zero, direct \CP violation in charged $B$ decays.
  To measure $R_{\CP\pm}$ and $A_{\CP\pm}$ we reconstruct \btodcppmk\ and
  \btodcppmp\ decays with the \dzcppm decaying to two \CP-odd and two \CP-even 
  eigenstates, and \btdk and \btdp decays with \Dz decaying to one non-\CP state.
  Previous measurements of these quantities were performed by \babar~\cite{prl_babar_d0cpeven}
  and Belle~\cite{bib:Belle_BtoDcpK}. 
  We update the result by \babar\ from $211$\invfb\ to $348$\invfb\ of data.
  Compared to the previous analysis, the current
  study does not include the decay mode $\Dz \ra \KS \phi$, since it
  is going to be explored by the Dalitz analysis method using $B^-{\to} DK^-$, $D{\to}
  K^0_S K^+ K^-$ decays. Dropping $\Dz \ra \KS \phi$ allows to combine the results
  of both measurements in the future.
  We also express the \CP-sensitive observables in terms of three
  Dalitz related independent quantities:
\begin{linenomath}
\begin{eqnarray}
  &&x_\pm=\frac{R_{\CP+}(1\mp A_{\CP+})-R_{\CP-}(1\mp A_{\CP-})}{4}\,,\\
  &&r^2=x_\pm^2+y_\pm^2=\frac{R_{\CP+}+R_{\CP-}-2}{2}\,,
\end{eqnarray}
\end{linenomath}
  where the Cartesian coordinates $x_\pm=r\cos(\delta\pm\gamma)$ and
  $y_\pm=r\sin(\delta\pm\gamma)$ are the same $\CP$ parameters as were 
  measured by the \babar\ Collaboration using $B^-{\to} DK^-, D{\to}
    K^0_S\pi^-\pi^+$ decays~\cite{babar_dalitz}.
  We reduce the systematic uncertainties from \Dz
  branching fractions and reconstruction efficiencies of different \Dz
  modes by measuring the double branching fraction ratios
\begin{linenomath}
\begin{equation}
  \label{eq:rpmdef}
    R_{\pm} = \frac{R^{K/\pi}_{\CP\pm}}{R^{K/\pi}}
\end{equation}
\end{linenomath}
  rather than the quantities $R_{\CP\pm}$. Here, 
\begin{linenomath}
\begin{equation}
    R^{K/\pi}_{\CP\pm} \equiv
    \frac{\BR(\bmtodcpk)+\BR(\bptodcpk)}{\BR(\bmtodcpp)+\BR(\bptodcpp)}\ ,
    \label{eq:def_Rkpi_cp}
\end{equation}
\end{linenomath} 
  and
\begin{linenomath}
\begin{equation}
    R^{K/\pi} \equiv
    \frac{\BR(\bmtodzk)+\BR(\bptodzbk)}{\BR(\bmtodzp)+\BR(\bptodzbp)}.
    \label{eq:def_Rkpi}
\end{equation}
\end{linenomath}
  $R_{\pm}$ and $R_{\CP\pm}$ are equivalent discarding a term of the order of $\approx 0.01$,
  which will be accounted for by assigning a systematic uncertainty when quoting the
  result in terms of $R_{\CP\pm}$.
  \footnote{The double branching fraction ratios, in the approximation
  $A(B^+ \to D^0_{{\rm CP}\pm}\pi^+) \approx A(B^- \to D^0_{{\rm
  CP}\pm}\pi^-) \approx \frac{1}{\sqrt 2}\,A(B^- \to D^0\pi^-)$, are
  equivalent to $R_{\CP\pm}$, discarding a term
  $r_B|V_{us}V_{cd}/V_{ud}V_{cs}|\approx 0.01$.}

\section{THE \babar\ DETECTOR AND DATASET}
  \label{sec:babar}
  This measurement uses 348\invfb\ of data taken at the 
  \FourS\ resonance by the \babar\ detector with the \pep2\ asymmetric $B$
  factory. 
  The \babar\ detector is described in detail
  elsewhere~\cite{detector}. 
  Tracking of charged particles is provided by a five-layer silicon
  vertex tracker (SVT) and a 40-layer drift chamber (DCH). A ring-imaging Cherenkov
  detector (DIRC) provides improved particle identification (PID). An electromagnetic
  calorimeter (EMC), comprised of CsI crystals,
  is used to identify electrons and photons. 
  These systems are mounted inside a $1.5$~T solenoidal
  magnetic field. The instrumented flux return of the 
  magnet allows discrimination of muons from other particles.
  We use a GEANT4-based Monte Carlo (MC) simulation~\cite{geant4} to model 
  the response of the detector, taking into account the varying
  accelerator and detector conditions. 

\section{EVENT SELECTION}
  \label{sec:Analysis}
  We reconstruct \btodh\ decays where the prompt track $h^-$ is a kaon
  or a pion. 
  Candidates for $D^0$ are
  reconstructed in the \CP-even eigenstates $\pi^-\pi^+$ and $K^-K^+$, in
  the \CP-odd eigenstates $\KS \pi^0$, $\KS \omega$
  and
  in the non-\CP flavor eigenstate $K^-\pi^+$. \KS and $\omega$ 
  candidates are selected in the \pipi and $\pi^+ \pi^- \pi^0$ 
  channels, respectively.

  PID information from the DCH and, when
  available, from the DIRC must be consistent with the kaon
  hypothesis for the $K$ meson candidate in all \Dz\ modes and with the pion
  hypothesis for the $\pi^\pm$ meson candidates in the $D^0{\ra}\pi^-\pi^+$
  mode. For the prompt track to be identified as a pion or a kaon,
  we require at least five Cherenkov photons to be detected
  to ensure a good measurement of the Cherenkov angle.
  We reject a candidate track
  if its Cherenkov angle is not within 4$\sigma$ of the expected value
  for either a kaon or pion mass hypothesis. We also reject
  candidate tracks that are identified as
  electrons by the DCH and the EMC 
  or as muons by the DCH and the muon system.

  Photon candidates are clusters in the EMC that are not matched to any
  charged track, have a raw energy greater than 30\mev and lateral
  shower shape consistent with the expected pattern of energy deposit
  from an electromagnetic shower.
  Photon pairs with invariant mass within the range 115--150\mevcc
  ($\sim$3$\sigma$) and
  total energy greater than 200\mev are considered \piz candidates.
  To improve the momentum resolution, the $\piz$ candidates are kinematically
  fit with their mass constrained to the nominal \piz\ mass~\cite{PDG2006}.

  Neutral kaons are reconstructed from pairs of oppositely charged
  tracks with invariant mass within 10\mevcc\
  from the nominal \KS mass~\cite{PDG2006}. We also require
  the ratio between the flight distance in the plane transverse to
  the beam direction and its expected uncertainty to be greater than 2.

  The invariant mass of a \Dz candidate
  must agree within $2.5\sigma$ of its mass resolution to the 
  nominal \Dz mass~\cite{PDG2006}. The \Dz mass resolution
  is about 7.5~\mevcc in the $K\pi$, \KpKm and \pipi modes, and about
  21~\mevcc and 9~\mevcc 
  in the $\KS \pi^{0}$ and $\KS \omega$ 
  modes, respectively.
  Selected \Dz candidates are fit with a constraint to the nominal \Dz mass.

  We reconstruct $B$ meson candidates by combining a \Dz\ candidate
  with a charged track $h^-$. For the $K^-\pi^+$ mode, the charge of the
  track $h^-$ must match the one of the kaon from the $D^0$ decay.
  We select $B$ meson candidates by using two kinematically independent variables: 
  the beam-energy-substituted mass
  \begin{linenomath}
  \begin{displaymath}
  \mes = \sqrt{(E_i^{*2}/2 + \mathbf{p}_i\cdot\mathbf{p}_B)^2/E_i^2-p_B^2}
  \end{displaymath}
  \end{linenomath}
  and the energy difference
  \begin{linenomath}
  \begin{displaymath}
  \Delta E=E^*_B-E_i^*/2,
  \end{displaymath}
  \end{linenomath}
  where the subscripts $i$ and $B$ refer to the initial \epem\ system and the
  $B$ candidate, respectively, and the asterisk denotes the
  beams center-of-mass (CM) frame.
  The \mes\ distributions for \btodh\ signal events are Gaussian
  distributions centered at the $B$ mass with a width of $2.6
  \mevcc$, which does not depend on the decay mode or on the nature of
  the prompt track.
  In contrast, the \DeltaE\ distributions depend on the mass assigned to the
  prompt track. We evaluate $\Delta E$ with the kaon mass hypothesis
  so that the distributions are centered near zero
  for \btodk\ events and shifted on average by approximately 50\mevc 
  to the positive direction for \btodp\ events.
  The \DeltaE\ resolution depends on the momentum resolutions of the
  \Dz meson and the prompt track $h^-$, and is typically $16\mev$
  for all the \Dz\ decay modes. All $B$ candidates are required to have
  \mes within 2.5$\sigma$ of the mean value and \DeltaE in the
  range $-0.15<\Delta E<0.20\gev$.

  To reduce background from continuum production of light quarks, we construct
  a Fisher discriminant based on the following four quantities:
  (i)  The Legendre polynomials, 
  a set of momentum-weighted sums of the tracks and neutrals not associated with 
  the reconstructed candidate, i.e. coming from the rest of the event (ROE):
\begin{linenomath}
\begin{equation}
    \label{LiExpression}
    L_j~=~\sum_i^{\rm ROE} p^*_i~\times~|cos(\theta^*_i)|^j,
\end{equation}
\end{linenomath}
  where $\theta_i^*$ is the CM angle between ${\bf p}^*_i$ and the thrust
  axis $\hat T^B$ of the $B$ candidate.
  We have considered only the ${L_0, L_2}$ pair, since previous studies
  have shown that adding other $L_j$ to the set
  of discriminating variables does not improve the sensitivity. In particular we use the 
  ratio ${L_2}/{L_0}$;
  (ii) $R_2^{\rm ROE}$, the ratio of the Fox-Wolfram moments
  $H^{\rm  ROE}_2/H^{\rm ROE}_0$,
  computed using tracks and photons in the ROE.
  $H^{\rm ROE}_l$ is defined as~\cite{fox_wol}:
\begin{linenomath}
\begin{equation}
    H^{\rm ROE}_l\equiv\sum_{i,j}^{\rm ROE}\frac{|{\bf p}_i^*||{\bf p^*}_j|}{E^{*2}_{\rm vis}}P_l(\cos\theta^*_{ij}),
\end{equation}
\end{linenomath}
  where $P_l$ is a Legendre polynomial, $\theta_{ij}$ is the opening angle
  between ${\bf p}^*_i$ and ${\bf p}^*_j$,
  and $E^*_{\rm vis}$ is the total visible
  energy of the event.
  (iii) $|\cos({\bf p}^*_B,z)|$ is the cosine of the angle of the $B$
  candidate momentum with respect to the beam ($z$) axis. 
  (iv) $|\cos(\hat{T}^B,z)|$ is the cosine of the angle of the $B$
  candidate thrust axis with respect to the $z$ axis.
  A cut on the value of the Fisher discriminant rejects more than
  $75\%$ of the continuum background while retaining about $85\%$  of the
  signal in all modes.

  Another source of background is related to \BB events. Its main contributions come
  from the processes $B{\ra}D^{*}h$ ($h=\pi,K$) and $B^-{\ra}D^0\rho^-$
  mis-reconstructed as \btodh candidates.
  For $D^0\ra K^-K^+$, $D^0\ra \pi^-\pi^+$, $D^0\ra K^0_S \pi^0$ and $\Dz \ra \KS \omega$
  decays, there are peaking backgrounds caused by $B$ mesons decaying into the same
  final state particles.
  The peaking backgrounds have
  \DeltaE\ and \mes\ distributions similar to the
  signal. Their yields
  are estimated from the $\Dz$ invariant mass sideband data and
  are taken into acount in the fit.

  When reconstructing $B$ candidates, it is possible that more
  than one combination satisfies the selection criteria in the same
  event.
  In order to select only one candidate per event, we
  define a criterion that allows to identify the
  combination with the largest probability to be a true signal \btodh\ decay.
  The $D^0$ invariant mass and the energy-substituted mass are chosen as
  discriminating quantities in all the channels. When \Dz decays into the \CP-odd
  channels we also include 
  the invariant masses of the $\omega$ and \piz candidates.
  These variables are combined in a $\chi^2$ function whose minimization defines the
  best candidate choice.
  In the end,
  the fraction of rejected background candidates in the selected samples is 2\% in the $K\pi$, less than 1\% for KK and \pipi
  modes, while it is about 5\% in the $\KS \pi^{0}$ mode and 8\% in
  the $\KS \omega$.
 
  The total reconstruction efficiencies, based on simulated
  signal events, are about 35\%($K^-\pi^+$), 32\%($K^-K^+$),
  33\%($\pi^-\pi^+$), 20\%($K^0_S\pi^0$)
  and 8\%($K^0_S\omega$).

\section{FIT PROCEDURE}
\label{sec:fit}
  We determine the signal and background yields for each \Dz\ decay mode from a
  two-dimensional extended unbinned maximum-likelihood fit
  to the selected data events determines the signal and background yields.
  The input variables to the fit are \DeltaE\ and a particle identification
  probability for the prompt track based
  on the Cherenkov angle $\theta_C$, the momentum $p$ and
  the polar angle $\theta$ of the track.
  The extended likelihood function $\mathcal{L}$ for the selected sample is given by the
  product of the probabilities for each individual candidate and a
  Poisson factor:
\begin{linenomath}
\begin{equation}\label{eq:ext_like}
  \mathcal{L}=\frac{e^{-N'}(N')^N}{N!}\prod_{i=1}^{N}\mathcal{P}_i.
\end{equation}
\end{linenomath}
  The probability $\mathcal{P}_i$ for a candidate in the event $i$ is the sum of the signal and background terms:
  \begin{linenomath}
  \begin{eqnarray}\label{eq:pdf}
  \mathcal{P}_i(\Delta E,\theta_C) &=& \frac{N_{\Dz\pi}}{N'}\ \mathcal{P}_i^{\Dz\pi}+\frac{N_{\Dz K}}{N'}\ \mathcal{P}_i^{\Dz\ K}+ \\
  &&\frac{N_{q\bar{q}(\pi)}}{N'}\ \mathcal{P}_i^{q\bar{q}(\pi)}+\frac{N_{q\bar{q}(K)}}{N'}\ \mathcal{P}_i^{q\bar{q}(K)}+\nonumber\\
  &&\frac{N_{B\bar{B}(\pi)}}{N'}\
  \mathcal{P}_i^{B\bar{B}(\pi)}+\frac{N_{B\bar{B}(K)}}{N'}\
  \mathcal{P}_i^{B\bar{B}(K)}+ \nonumber\\
  &&\frac{N_{X_1X_2K}}{N'}\ \mathcal{P}_i^{X_1X_2 K}. \nonumber
  \end{eqnarray}
  \end{linenomath}
  where
  $N'=N_{\Dz\pi}{+}N_{\Dz
  K}{+}N_{q\bar{q}(\pi)}{+}N_{q\bar{q}(K)}{+}N_{B\bar{B}(\pi)}{+}N_{B\bar{B}(K)}{+}N_{X_1X_2K}$.
  Each addendum on the right-hand side of equation (\ref{eq:pdf}) is
  the product of two different terms. The ratio $N_J/N'$
  ($J=D^0\pi, D^0K,...$) represents the probability to choose a
  candidate of type $J$ after the selection criteria are applied; the
  term $\mathcal{P}^J_i$ is the probability to measure the particular
  set of physical quantities $\{\DeltaE,\theta_C$\}$_i$ in the $i^{th}$
  event, once the candidate of type $J$ has been selected:
  \begin{linenomath}
  \begin{equation}\label{eq:pdf_like}
  \mathcal{P}_i^J=\mathcal{P}^J_{\Delta E,i}\ \mathcal{P}^J_{\theta_C,i}.
  \end{equation}
  \end{linenomath}
  
  The \DeltaE\ distribution for \btdk\ signal is parameterized with a double Gaussian 
  function. The mean and width of the narrow Gaussian are denoted in the following with $\mu(\Dz K)$ and
  $\sigma(\Dz K)$. The \btdp\ \DeltaE probability density function (PDF) would be the same as the \btdk\ one, if the
  prompt track had been assigned the pion mass. Since \DeltaE is
  computed by assigning the kaon mass, it is shifted by a quantity
  \begin{linenomath}
  \[\DeltaE_{\rm shift}(\gamma,|\vec{p}|)=\gamma\left(\sqrt{m_K^2+|\vec{p}|^2} -
  \sqrt{m_{\pi}^2+|\vec{p}|^2}\right)\]
  \end{linenomath}
  which depends on the momentum $\vec{p}$ of the prompt track in the
  lab frame. $\gamma$ is the Lorentz parameter characterizing the boost of the
  center of mass frame relative to the lab frame. Therefore we parameterize 
  the \btdp \DeltaE shape with
  a double Gaussian whose mean is computed, event-by-event, as $\mu(\Dz
  \pi) = \mu(\Dz K) + \DeltaE_{\rm shift}(\gamma,|\vec{p}|)$, and
  whose width is the same as for the \btdk signal component.
  The fraction of the wide component of the signal shape,
  its offset from the narrow component and the ratio between its width
  and the width of the narrow component are fixed using the mode-dependent 
  numbers obtained from the MC simulation. 
  The \DeltaE\ distributions for the continuum background 
  are parameterized with a first order polynomial.
  The \DeltaE\ distribution for the \BB background is empirically
  parametrized with a ``Crystal-Ball'' lineshape~\cite{crystal_ball}: a Gaussian with an exponential 
  tail at higher  \DeltaE values. The parameters of the background shapes 
  are determined from MC simulated events and are fixed in the fit.

  The particle identification PDF is obtained from MC simulation.
  Its parametrization is performed
  by means of a double Gaussian distribution as a function of $\theta_C^{\rm pull}$,
  which is the difference between the measured Cherenkov angle
  and its expected value for a given mass hypothesis, divided by the expected error.

  We independently fit five samples corresponding to each of the five $\Dz$ decay modes under study. 
  The fit simultaneously evaluates separate likelihood functions
  for \Bp and \Bm categories. 
  In the fit the free parameters 
  are $\Dz K$ and $\Dz \pi$ signal yield asymmetries, total number of signal 
  events in $\Dz \pi$ ($N_{\Dz \pi}$), ratio $R_{K/\pi}$ = $N_{\Dz K}/N_{\Dz \pi}$, eight background
  yields: $N_{\qqbar(\pi)}$, $N_{\qqbar(K)}$,
  $N_{\BB(\pi)}$, $N_{\BB(K)}$ (one for each charge, i.e. $4 \times 2 = 8$), 
  and two parameters of the \DeltaE
  signal shape (shared between positive and negative samples).
  The number of peaking
  background events $N_{X_1X_2 K}$ is fixed to the values obtained from the
  study using $\Dz$ mass sidebands.
  We assume no charge asymmetry in
  the peaking background (a small systematic error due to this assumption 
  is considered later).

\section{PHYSICS RESULTS AND SYSTEMATIC STUDIES}
  \label{sec:Systematics}
\begin{figure}[p]
  \begin{center}
    \includegraphics[height=4.0cm,width=0.42\textwidth,angle=0]{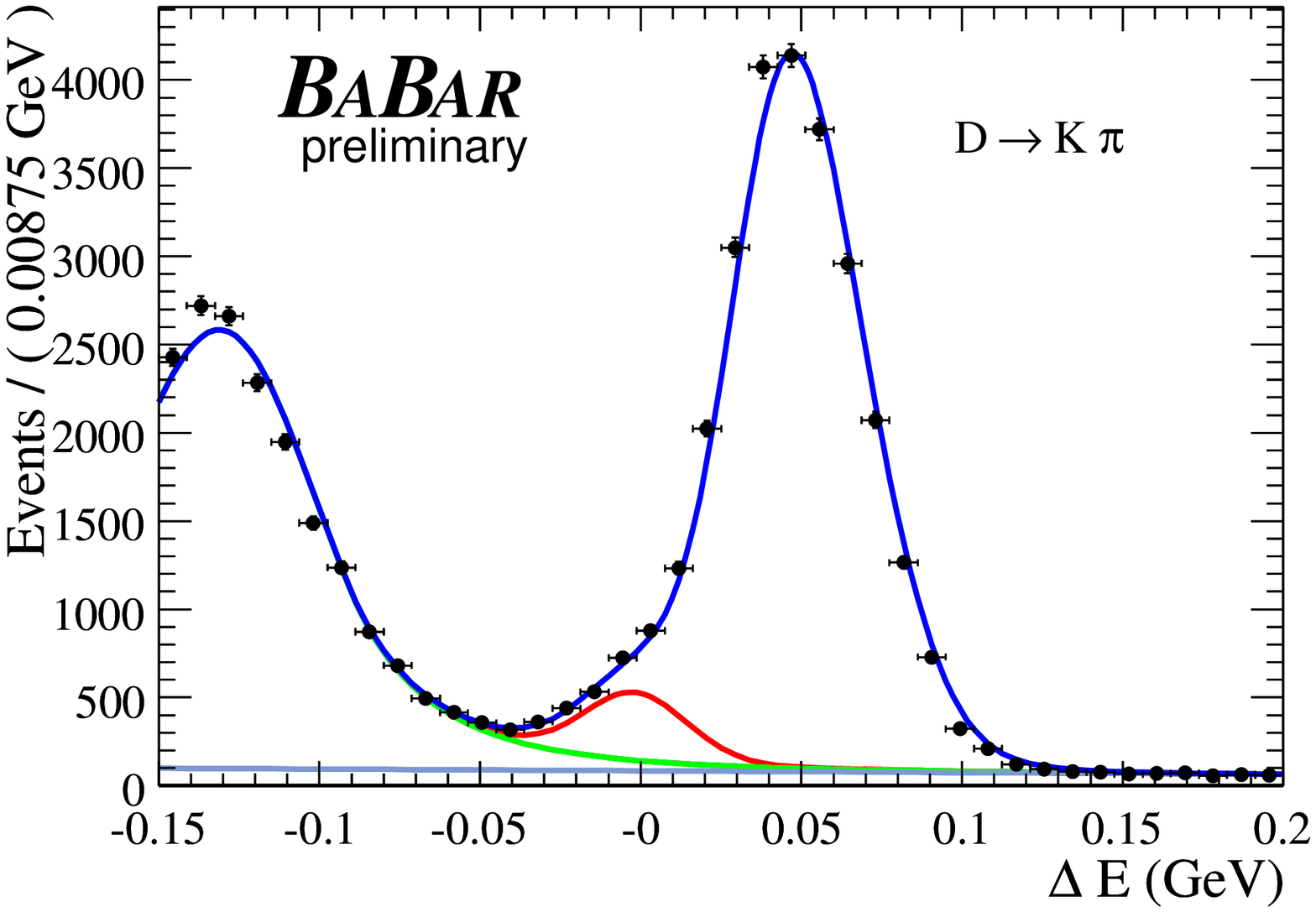}
    \includegraphics[height=4.0cm,width=0.42\textwidth,angle=0]{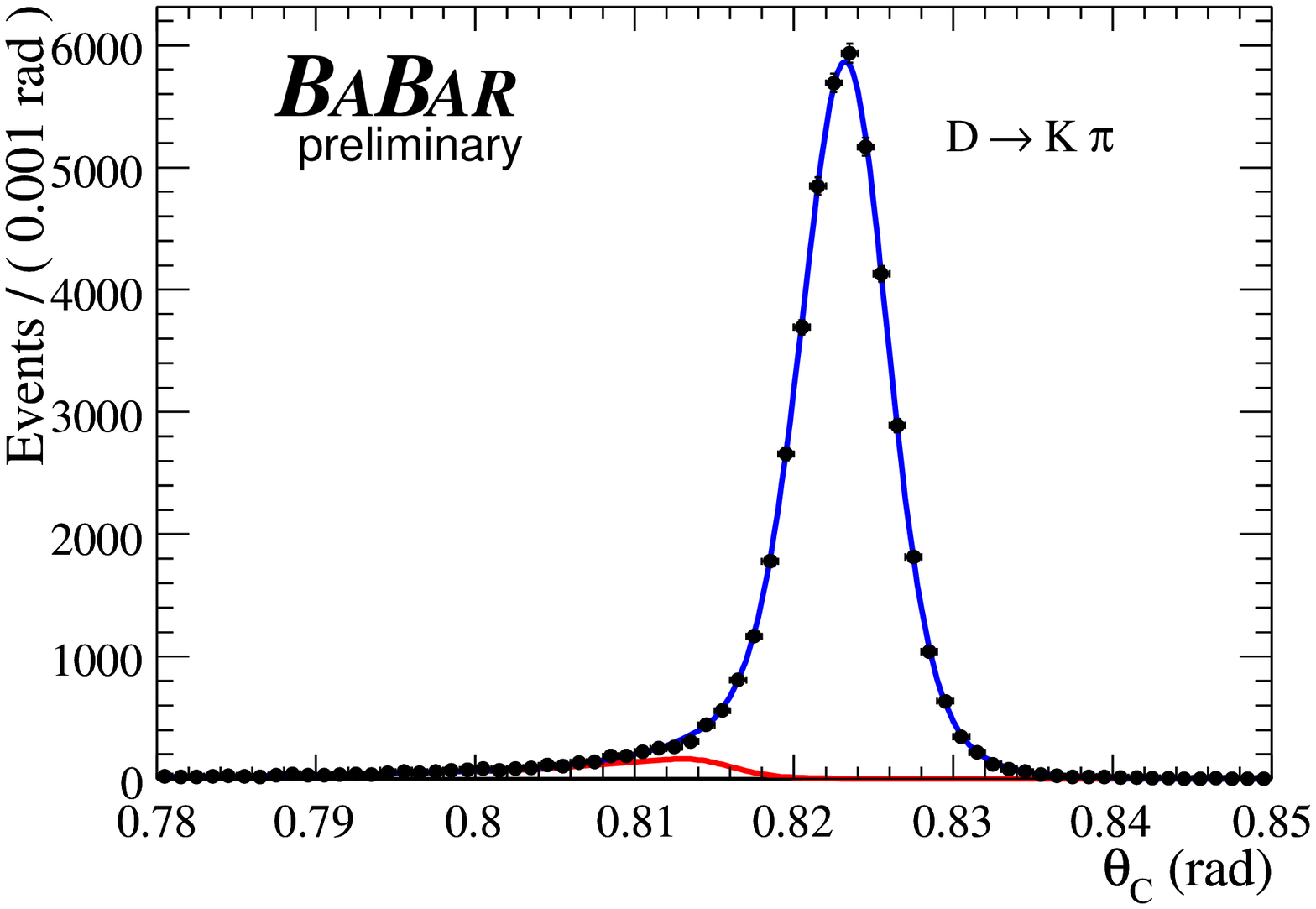}
    \includegraphics[height=4.0cm,width=0.42\textwidth,angle=0]{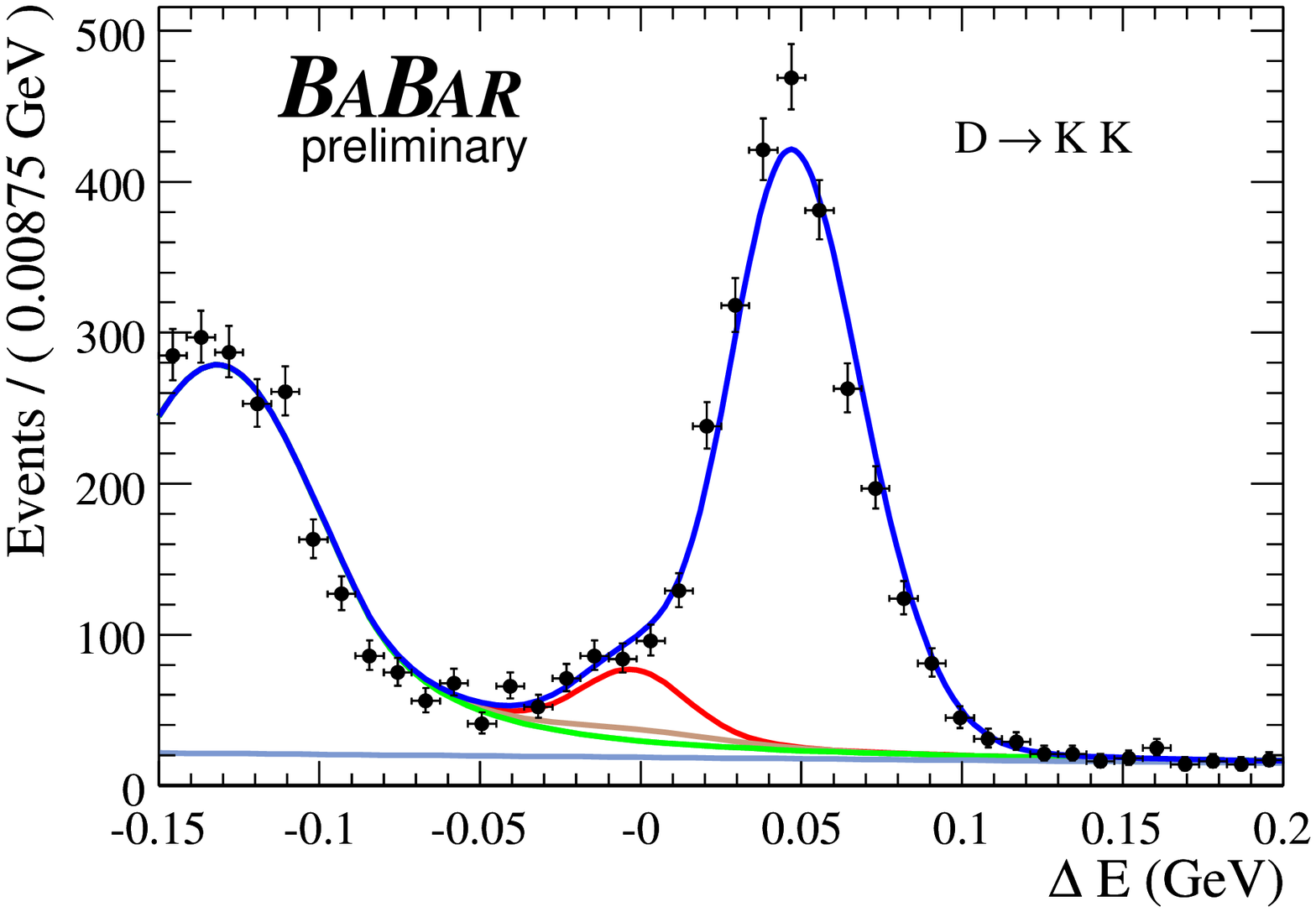}
    \includegraphics[height=4.0cm,width=0.42\textwidth,angle=0]{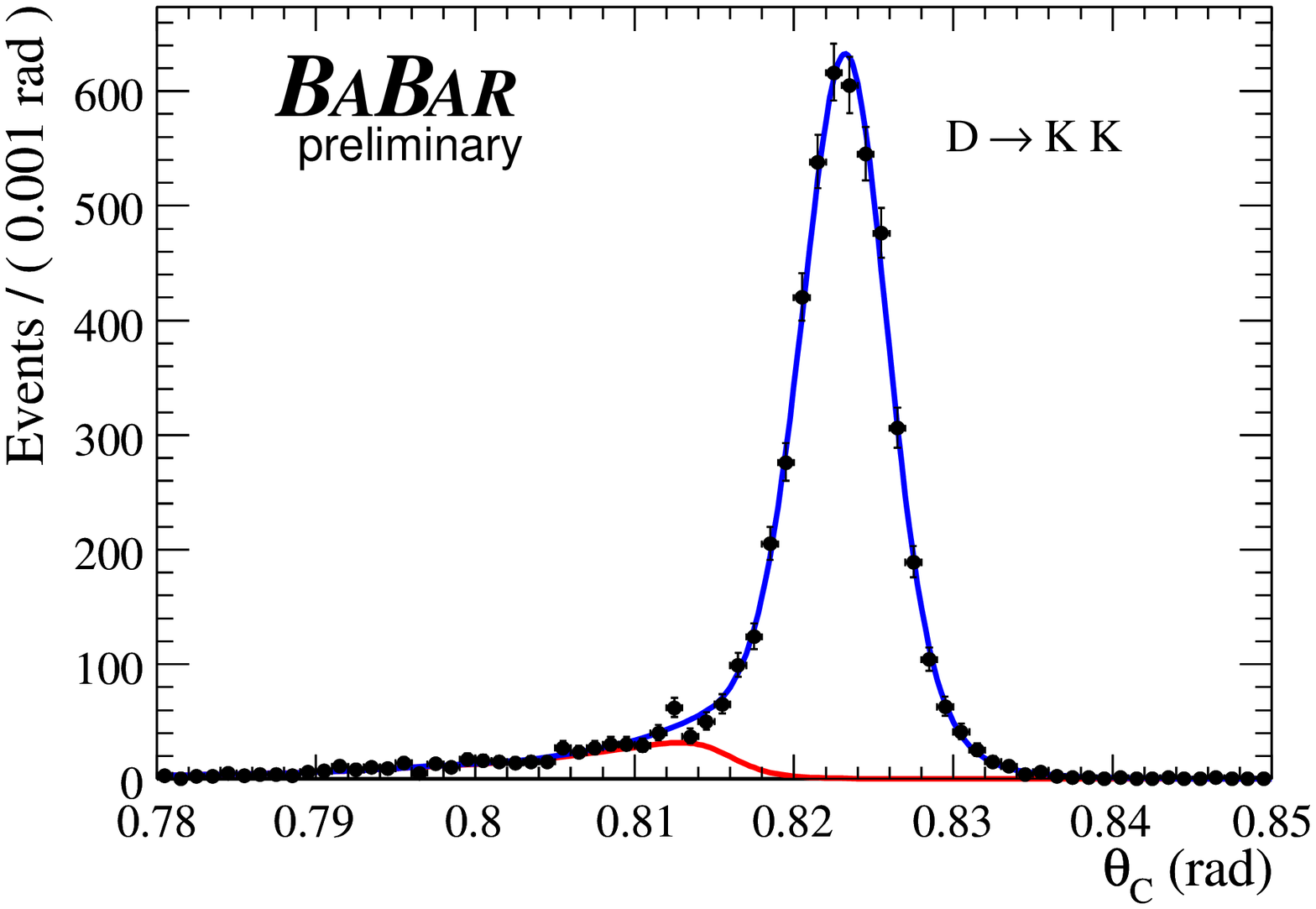}
    \includegraphics[height=4.0cm,width=0.42\textwidth,angle=0]{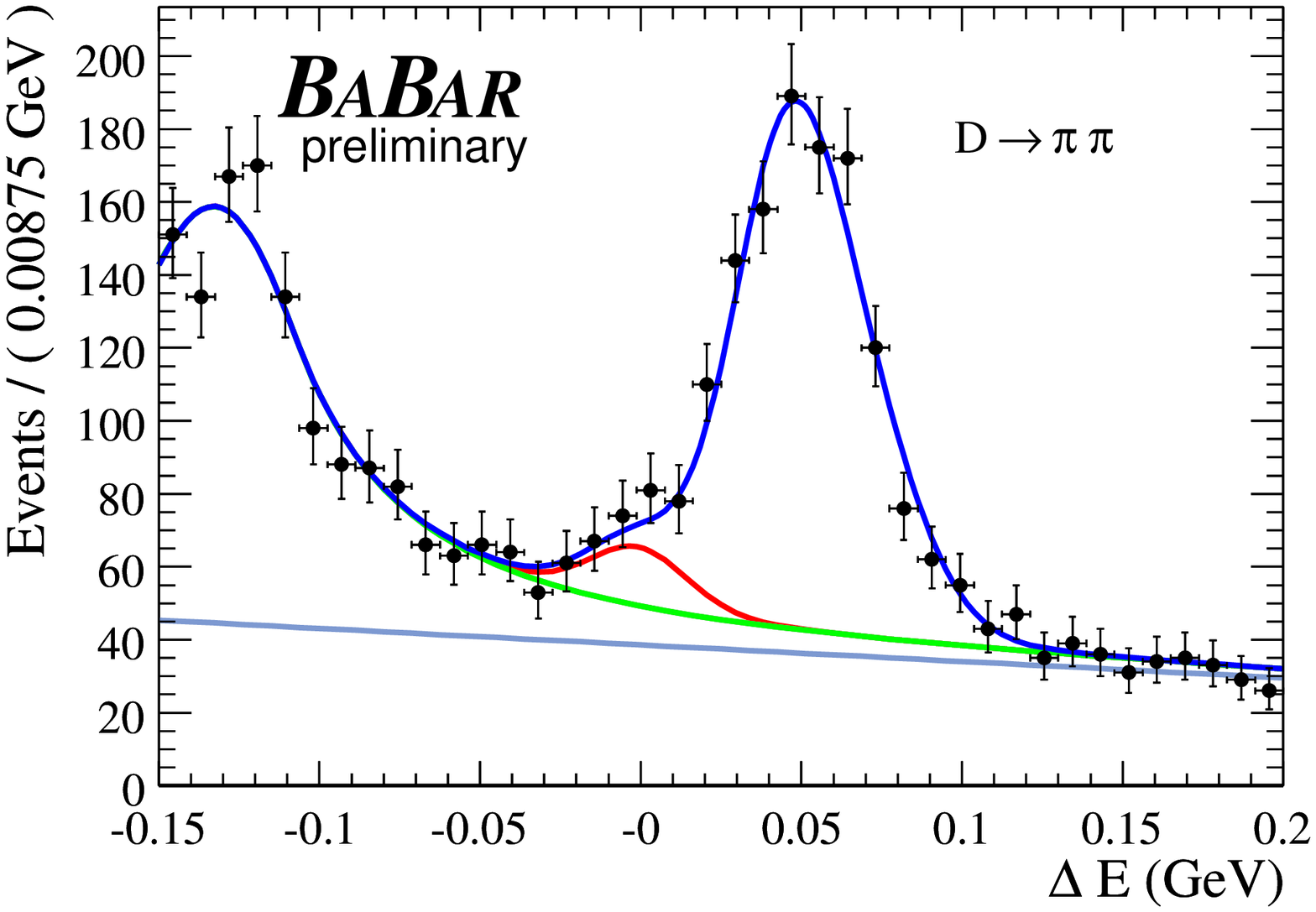}
    \includegraphics[height=4.0cm,width=0.42\textwidth,angle=0]{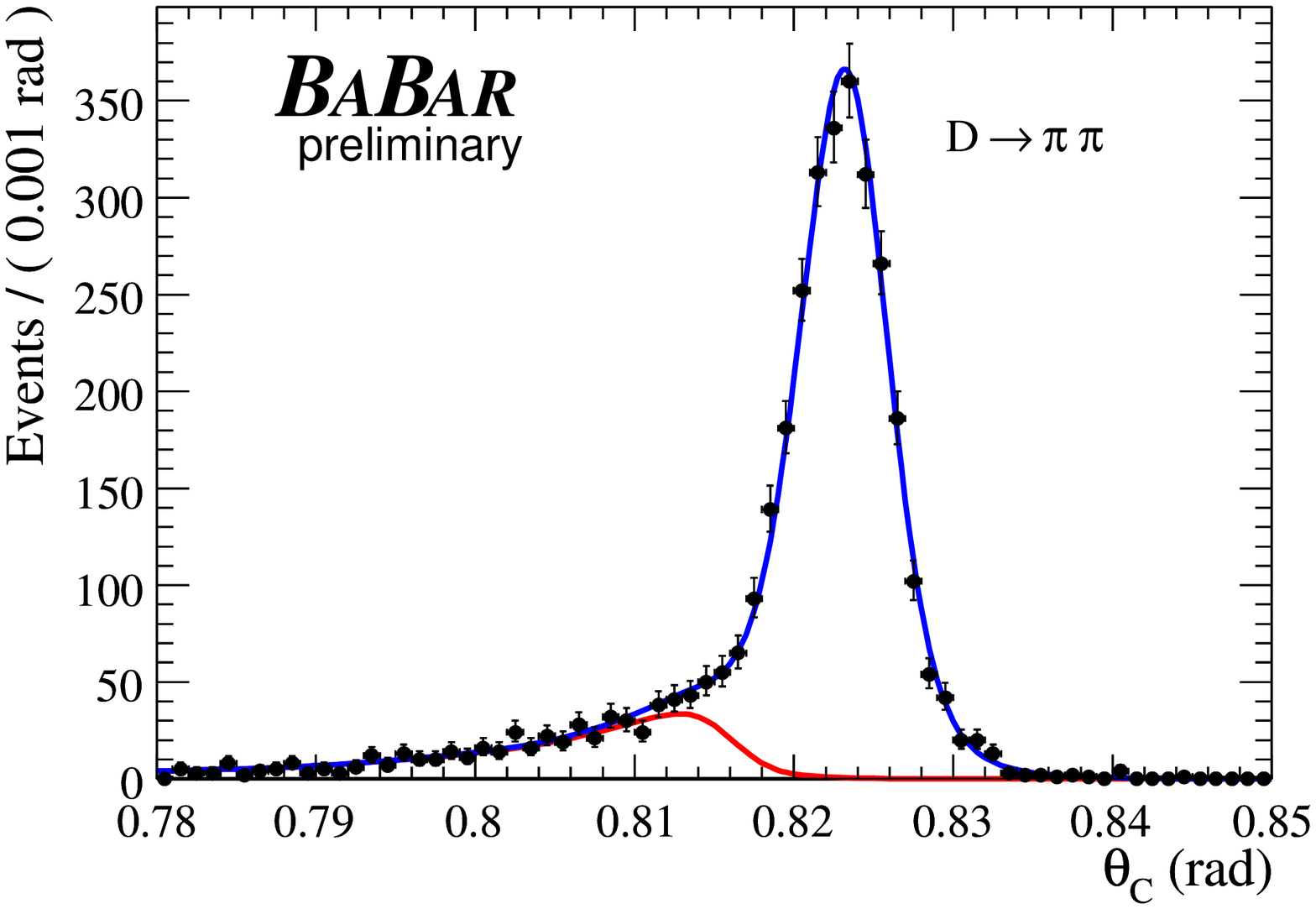}
    \includegraphics[height=4.0cm,width=0.42\textwidth,angle=0]{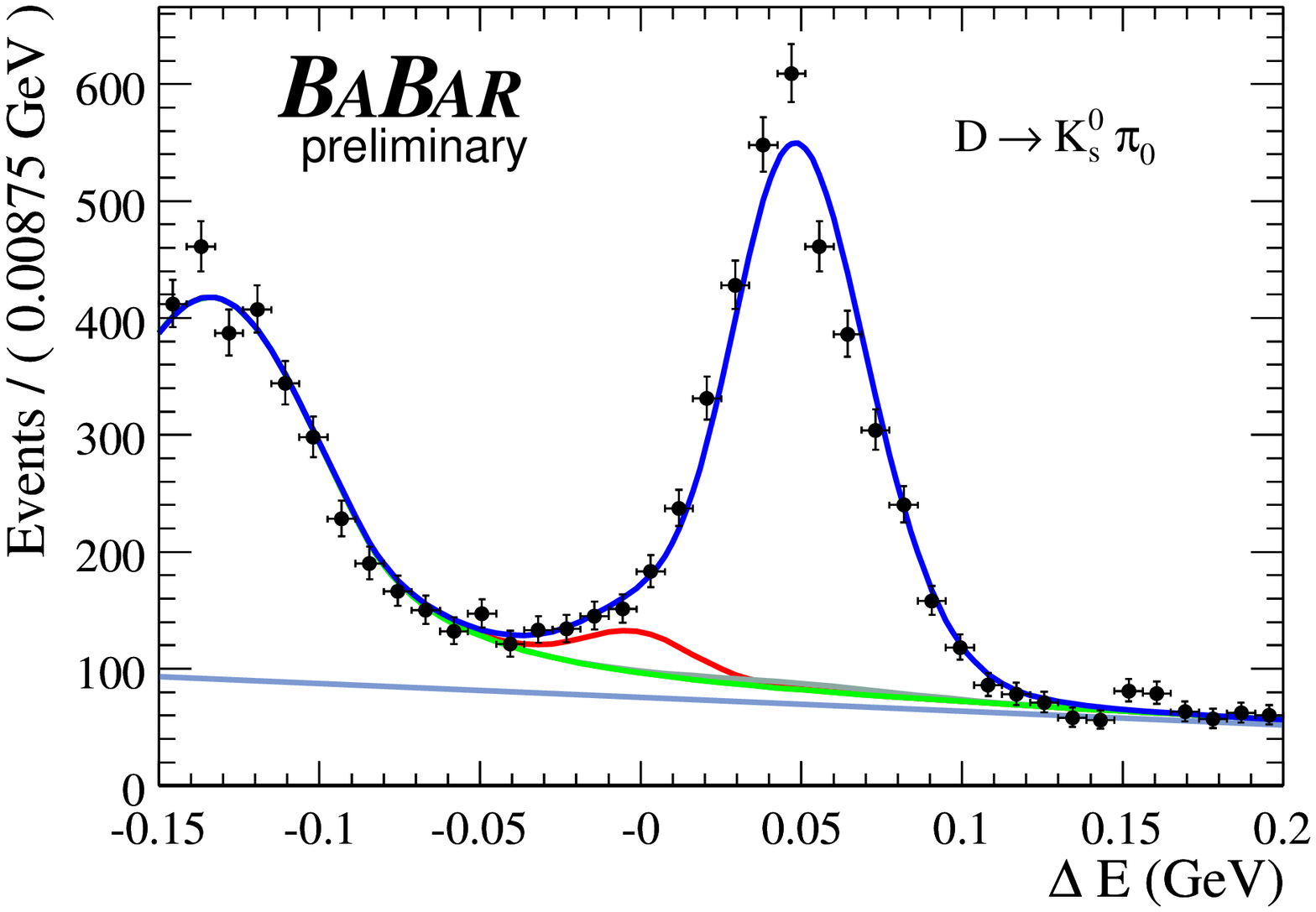}
    \includegraphics[height=4.0cm,width=0.42\textwidth,angle=0]{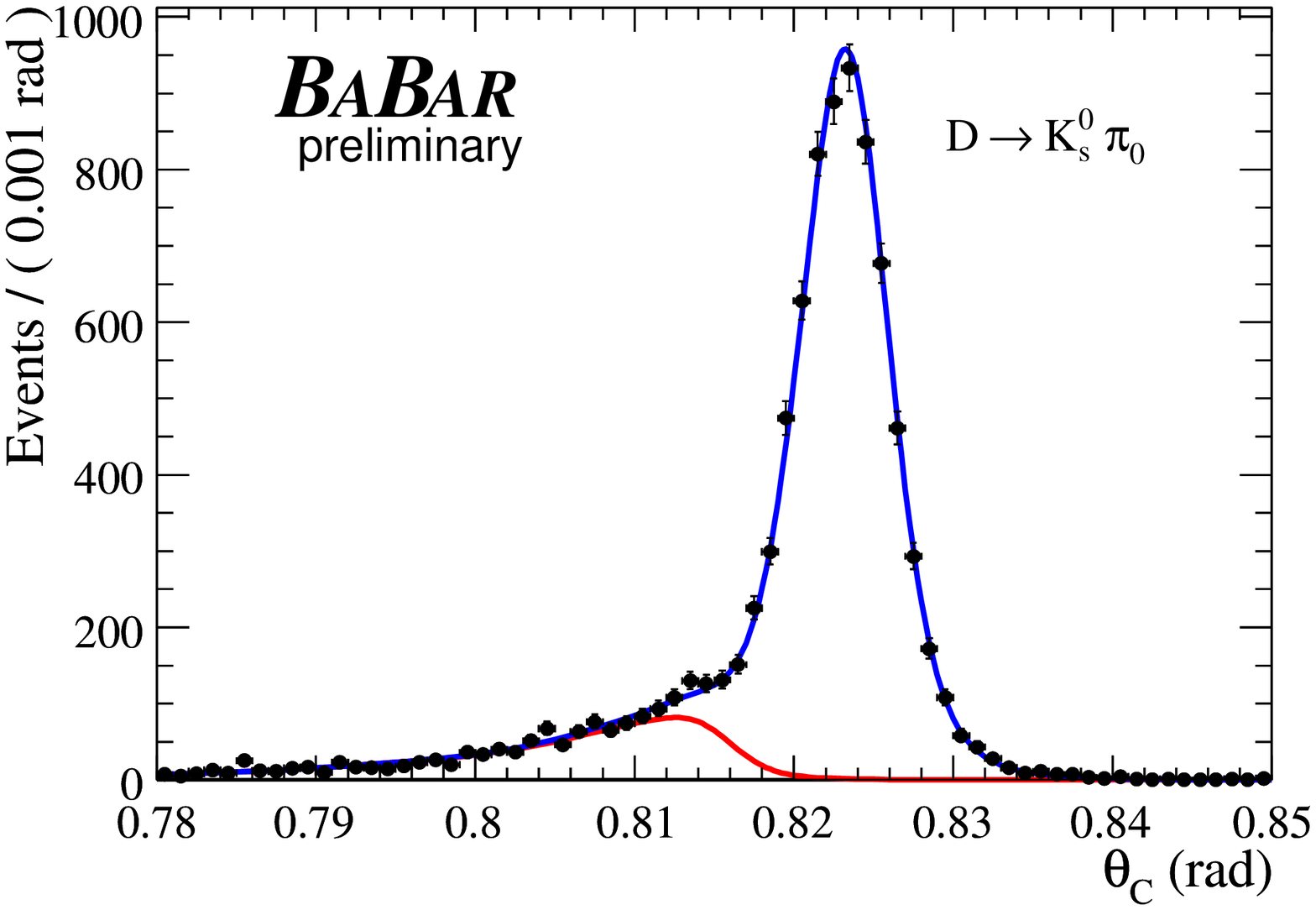}
    \includegraphics[height=4.0cm,width=0.42\textwidth,angle=0]{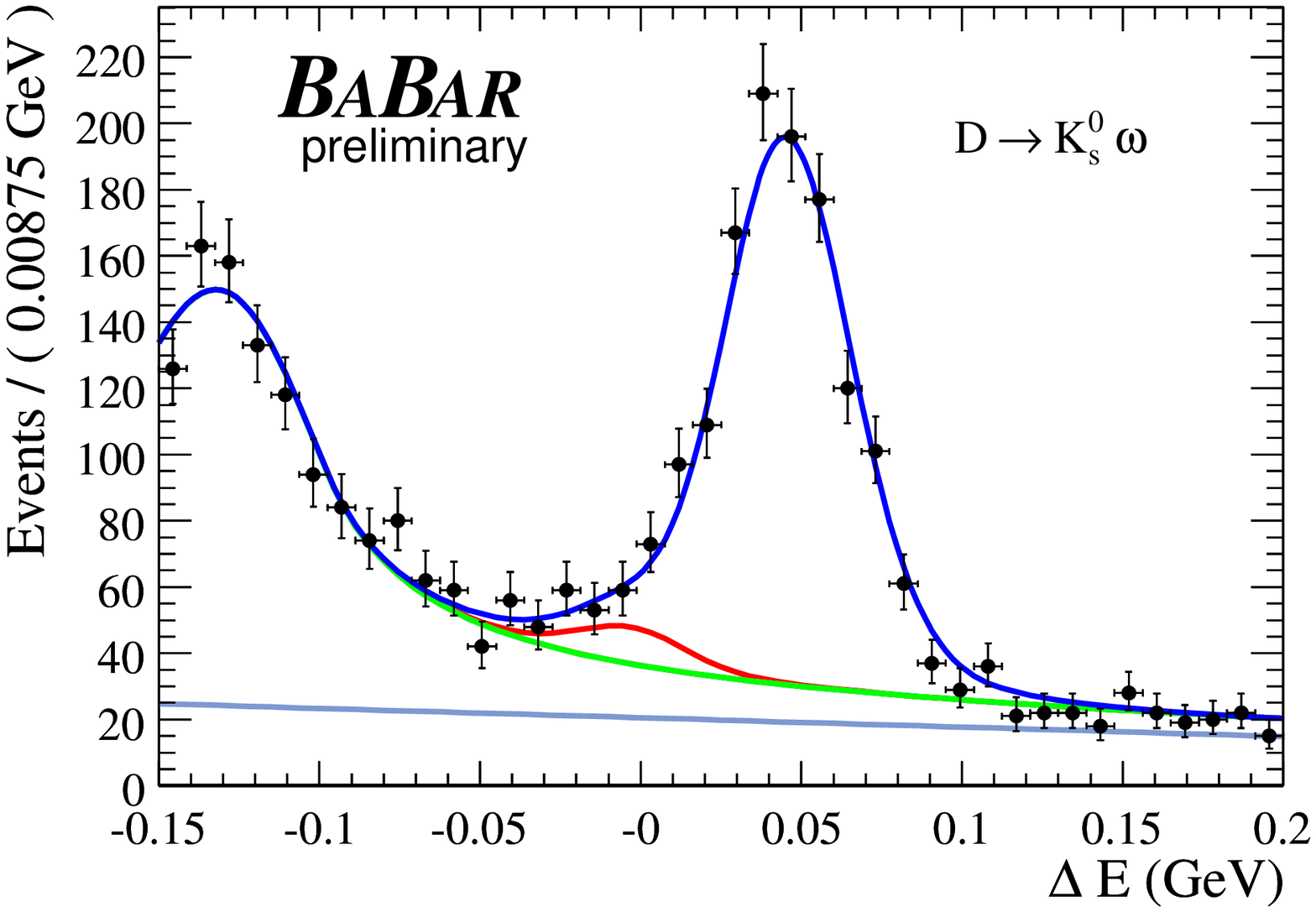}
    \includegraphics[height=4.0cm,width=0.42\textwidth,angle=0]{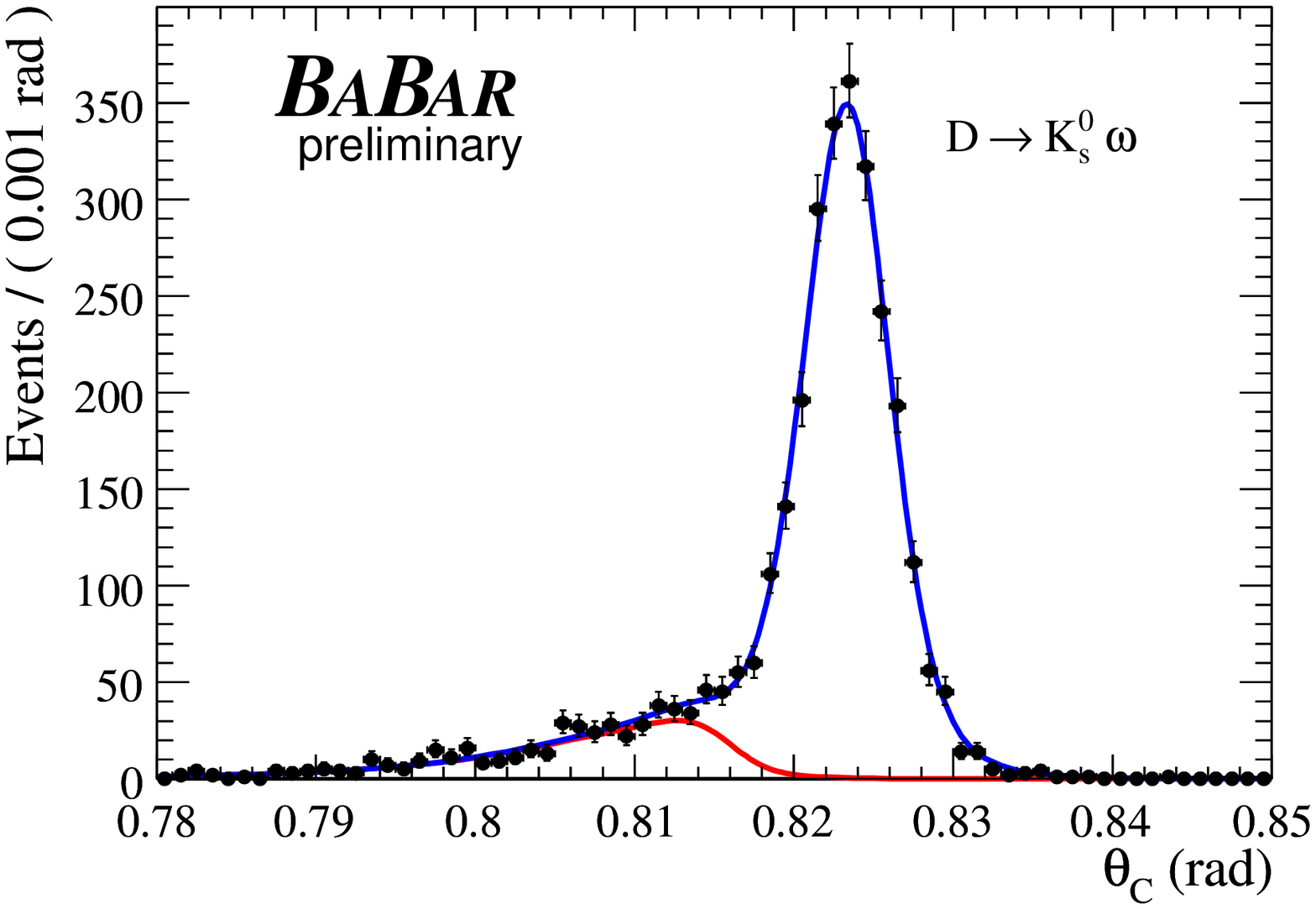}
    \caption{
    $\Delta E$ (left) and $\theta_C$ (right) distributions
    of selected $B^- \ra \Dz h^-$ events.
    The blue line represents the
    projection of the likelihood in the plotted variable.
    The red line represents the \btdk component.
    In the left hand plots, the green and light blue lines indicate 
    $B{\bar B}$ and continuum backgrounds, respectively. The brown line refers to the peaking background 
    (when present). 
    }
    \label{fig:fit}
  \end{center}
\end{figure}

\begin{figure}[p]
  \begin{center}
    \includegraphics[height=4.5cm,width=0.44\textwidth,angle=0]{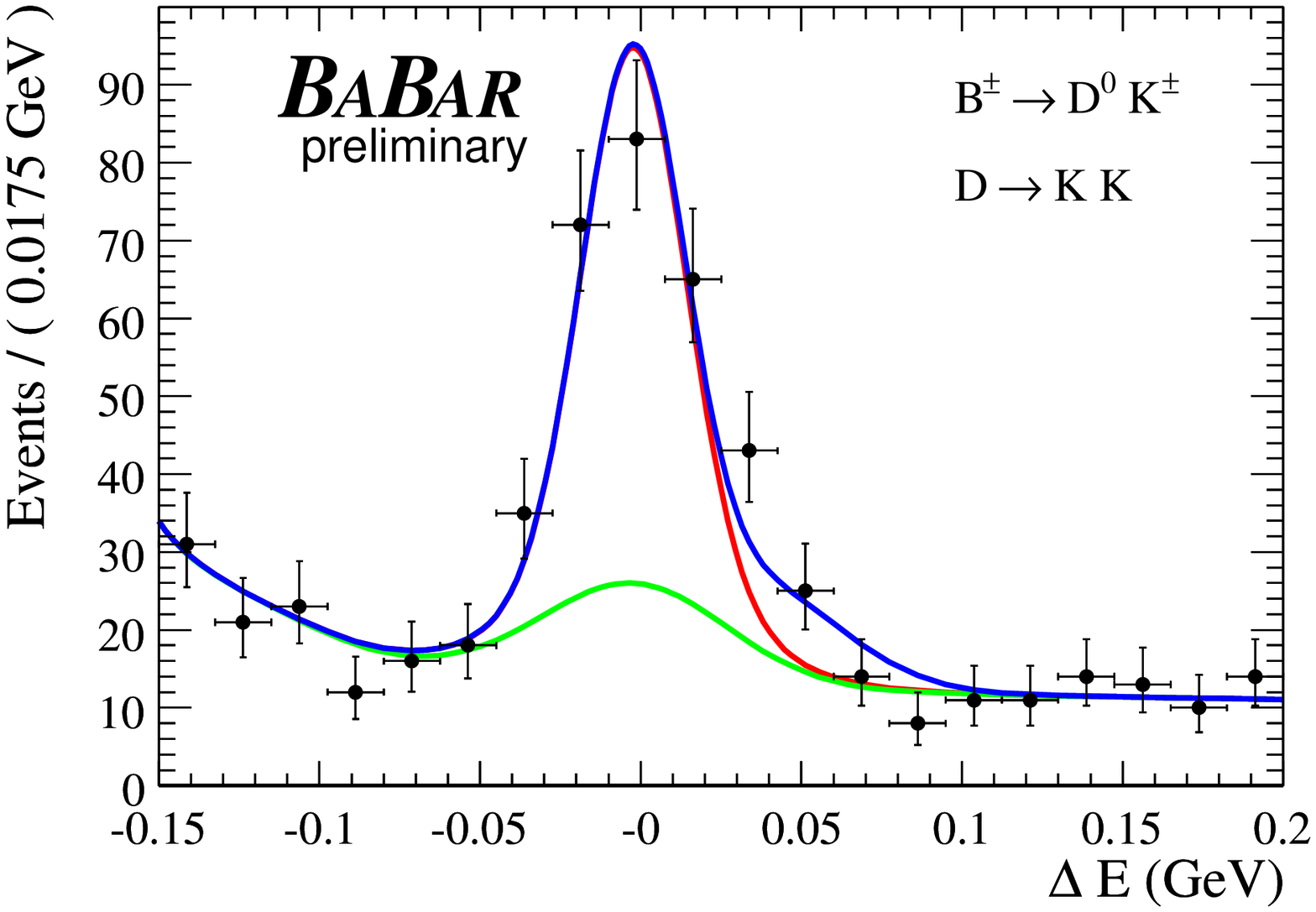}
    \includegraphics[height=4.5cm,width=0.44\textwidth,angle=0]{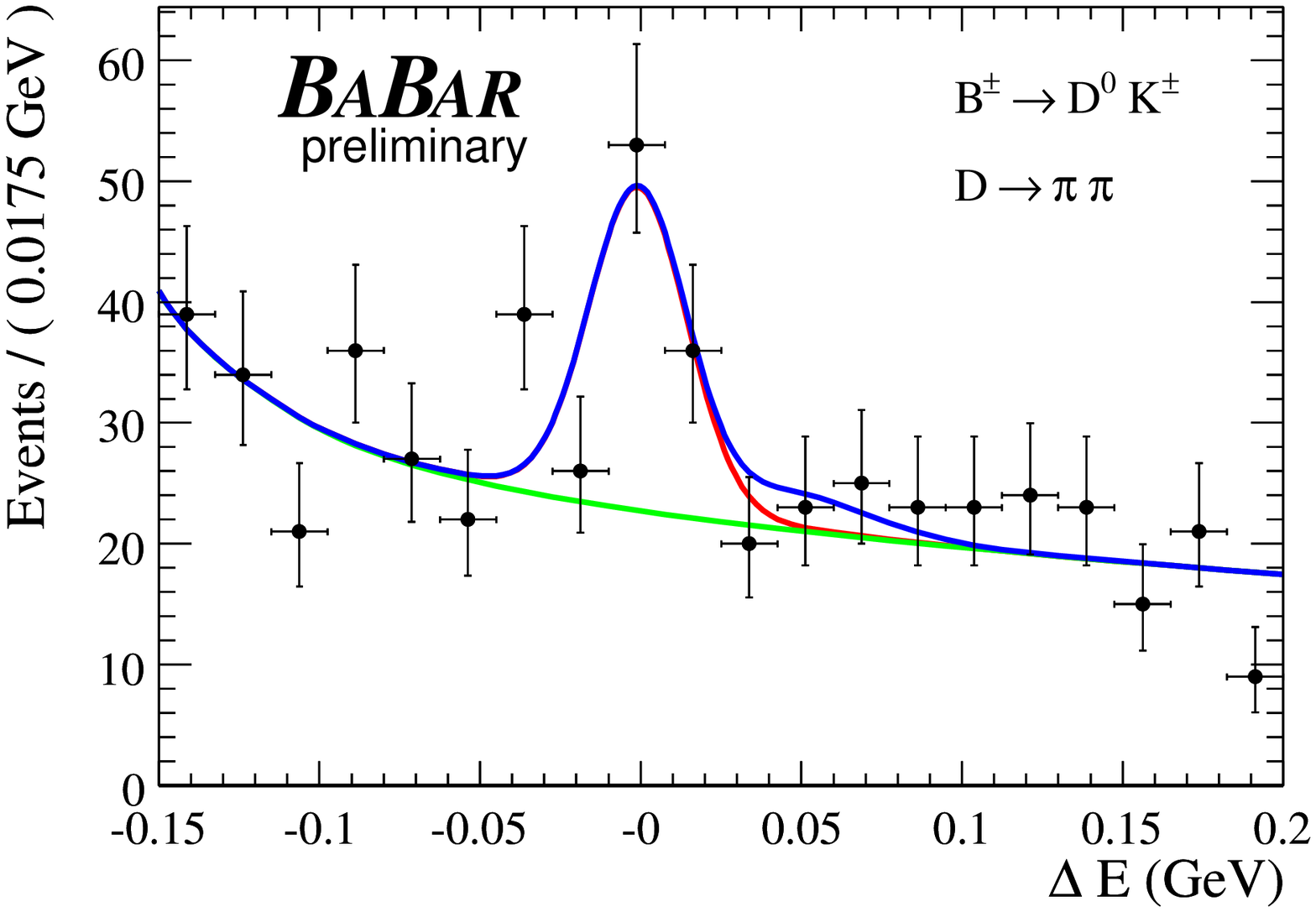}
    \includegraphics[height=4.5cm,width=0.44\textwidth,angle=0]{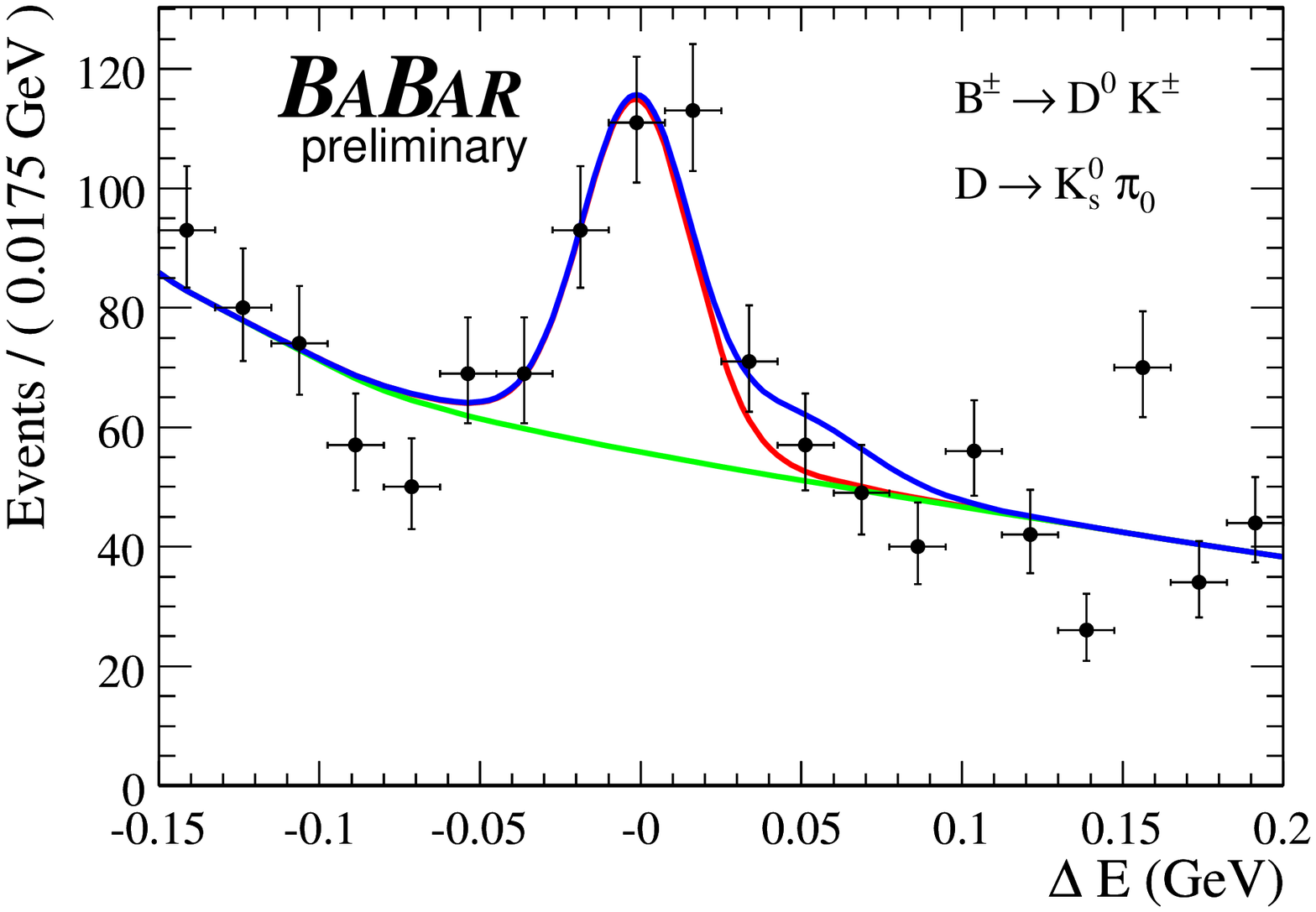}
    \includegraphics[height=4.5cm,width=0.44\textwidth,angle=0]{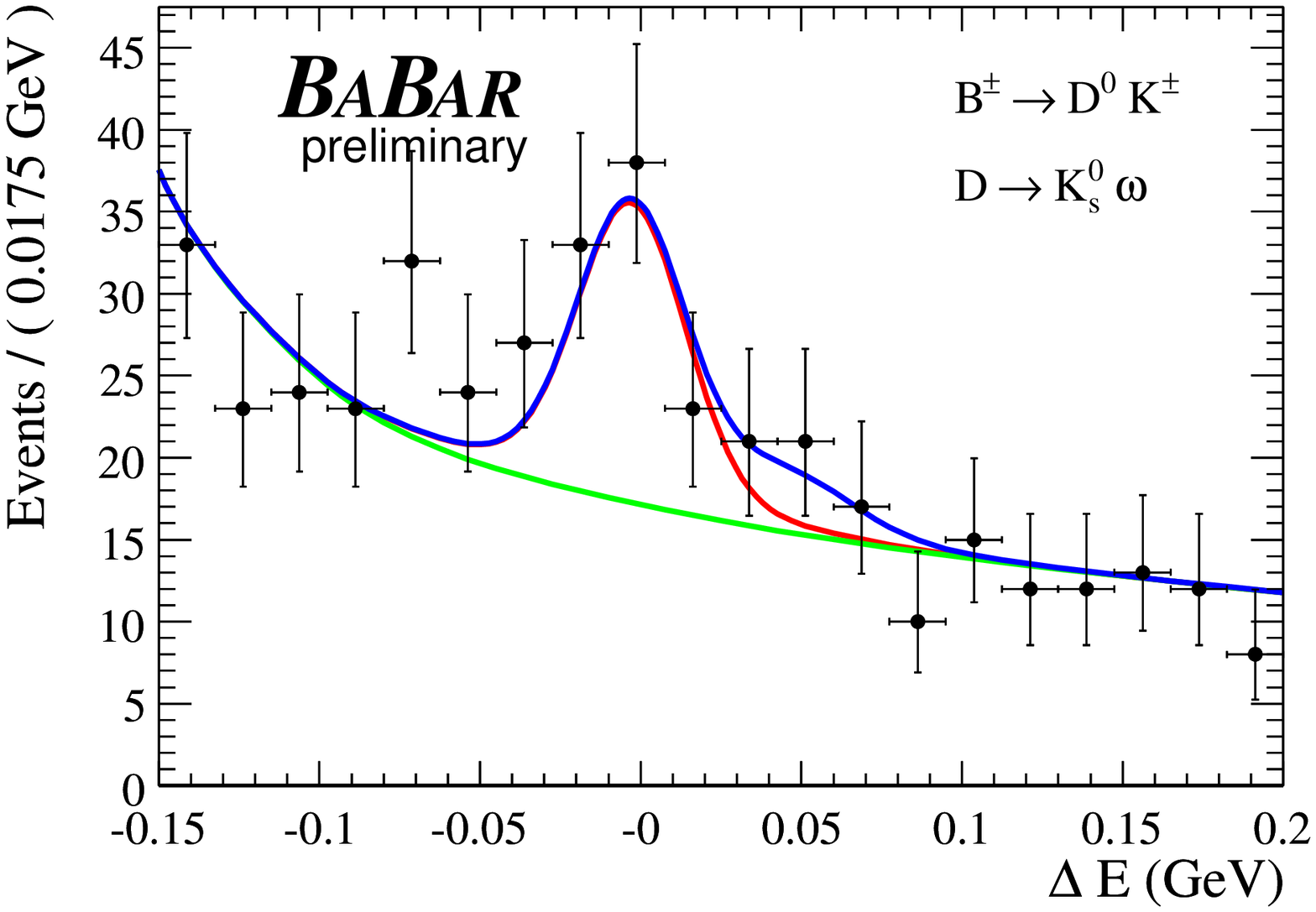}
    \includegraphics[height=4.5cm,width=0.44\textwidth,angle=0]{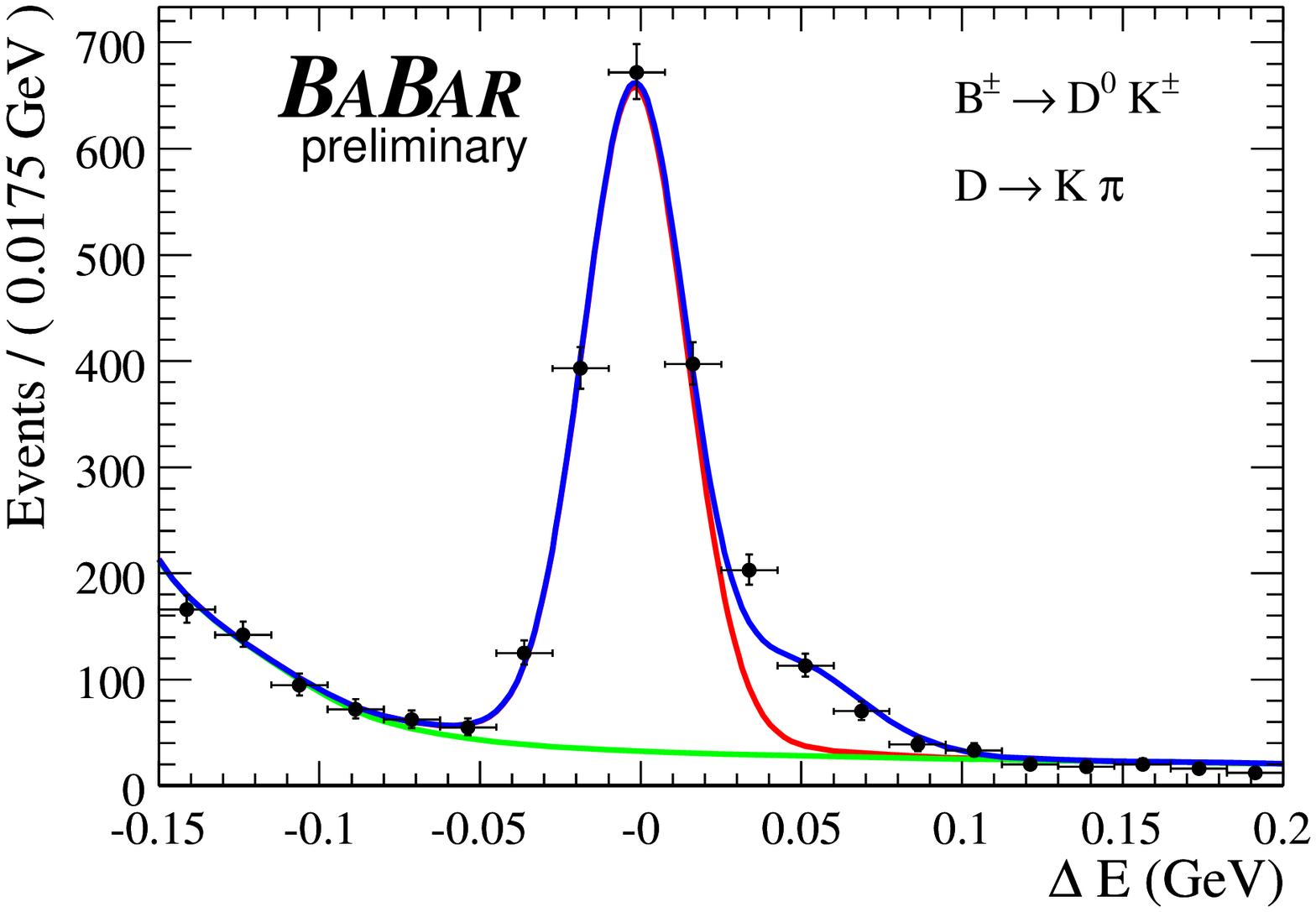}
    \caption{
    $\Delta E$ distributions
    of $B^- \ra \Dz K^-$ signal enhanced $B ^-\ra \Dz h^-$ events.
    The blue line represents the
    projection of the likelihood,
    the red line indicates the \btdk component, the green line shows the total background
    contribution.
    The remaining \btdp signal is visible as a small shoulder on the right hand side of
    the \btdk signal peak.
    }
    \label{fig:fit-sek}
  \end{center}
\end{figure}

  The results of the fit are summarized in Table~\ref{tab:fitresults}.
  Figure~\ref{fig:fit} shows the distributions of \DeltaE\ and $\theta_C$ for the 
  $K^-\pi^+$, \CP-even and \CP-odd modes.
  The projections of the likelihood fits
  are overlaid on the plots.
  On Figure~\ref{fig:fit-sek} we show the $\Delta E$ projections produced with a
  kaon identification requirement applied to the prompt track. Hence the
  $B^- \ra \Dz K^-$ signals become prominently visible on the plots, while $B^- \ra \Dz \pi^-$
  contributions significantly decrease.

\begin{table}[h]
    \caption{\btodk\ and \btodp\ signal event yields obtained from the fit to the data.
    All values are preliminary.}
    \label{tab:fitresults}
    \begin{center}
      \begin{tabular}{lcccc}
        \hline
        \hline
        $D^0$ mode &\  $N(B\ra D^0\pi)$ &\ $N(B\ra D^0K)$ & $N(B^+\to \Dz K^+)$ & $N(B^-\to\Dzb K^-)$\\
        \hline
        $K^-\pi^+$      &\  $24965\pm 169$ &\ $1859\pm 52$ & $951\pm 36$ & $909\pm 35$\\ 
        \hline
        $K^-K^+$        &\  $ 2412\pm  54$ &\ $ 189 \pm 20$ & $61 \pm 11$ & $128 \pm 16$\\
        $\pi^-\pi^+$    &\  $  876\pm  38$ &\ $ 73 \pm 15$ & $24\pm 9$ & $49 \pm 12$ \\
        \hline
        $K^0_S\pi^0$    &\  $ 2967 \pm 69$ &\ $ 184 \pm 25$ & $107 \pm 19$ & $77 \pm 16$ \\
        $K^0_S\omega$    &\  $ 1023 \pm 44$ &\ $ 59 \pm 14$ & $ 36 \pm 12$ & $ 23 \pm 9$ \\
        \hline
        \hline
      \end{tabular}
    \end{center}

  \end{table}

  The double ratios $R_{\CP\pm}$ are computed by calculating a weighted mean of the ratios $R_{K/\pi}$
  for \CP-even and \CP-odd modes and dividing it by $R_{K/\pi}$ for the non-\CP\ mode.
  Correction factors (ranging from 1.006 to 1.027 depending on the
  \Dz mode) that account for small differences in the efficiency 
  between the \btodk\ and \btodp\ selections are taken into account. An additional factor is 
  applied to the results in the $\Dz \ra \KS \omega$
  mode to correct for a dilution due to the S-wave non-resonant
  contribution. These corrections were estimated using a fit to
  the $\omega$ helicity angle in the selected data events and found to
  be $1.10 \pm 0.11$ for $A_{\CP-}^{\KS \omega}$ and $0.98 \pm 0.02$ for $R_{\CP-}^{\KS \omega}$. 
  The uncertainties in the correction factors
  are included in the systematic errors ($\pm 0.006$ and $\pm 0.008$ for $R_{\CP-}$
  and $A_{\CP-}$, respectively).
  The results for each mode separately and combined by \CP-even and \CP-odd 
  categories are listed in Table~\ref{tab:final_ratio}.

  \begin{table}[!h]
    \caption{Measured double branching fraction ratios $R_{\CP\pm}$ and \CP
      asymmetries $A_{\CP\pm}$ for different \Dz\ decay modes. In the combined results, the first
      error is statistical, the second is systematic. For individual modes, only statistical
      errors are shown. All values are preliminary.} 
    \label{tab:final_ratio}
    \begin{center}
      \begin{tabular}{lcc}
        \hline
        \hline
        $D^0$ decay mode &\ \ \ \ \ \ $R_{\CP}$ &\ \ \ \ \ \ $A_{\CP}$\\
        \hline
        \hline
        $K^-K^+$ &\ \ \ \ \ \  $1.05\pm 0.11$&\ \ \ \ \ \ $0.36\pm 0.10$ \\
        $\pi^-\pi^+$ &\ \ \ \ \ \  $1.13\pm 0.22$&\ \ \ \ \ \ $0.33\pm 0.20$ \\
        \CP-even combined &\ \ \ \ \ \ $1.07\pm0.10\pm0.04$&\ \ \ \ \ \ $0.35\pm 0.09\pm 0.05$\\
        \hline
        $\KS\pi^0$ &\ \ \ \ \ \  $0.84\pm 0.11$&\ \ \ \ \ \ $-0.16\pm 0.13$ \\
        $\KS\omega$ &\ \ \ \ \ \  $0.75\pm 0.18$&\ \ \ \ \ \ $-0.24\pm 0.26$ \\
        \CP-odd combined &\ \ \ \ \ \ $0.81\pm0.10\pm0.05$&\ \ \ \ \ \ $-0.19\pm 0.12\pm 0.02$\\
        \hline
        \hline
      \end{tabular}
    \end{center}
  \end{table}
 
  Systematic uncertainties in the double ratios $R_{\CP\pm}$ and in the \CP 
  asymmetries $A_{\CP\pm}$ arise primarily from uncertainties in
  signal yields due to the estimate of the peaking backgrounds 
  ($\pm 0.03$ for $R_{\CP+}$ and $\pm 0.05$ for $R_{\CP-}$) and
  from the imperfect knowledge of the PDF shapes.
  The parameters of the PDFs that are fixed in the nominal fit are
  varied by $\pm1\sigma$ and the observed difference in the parameters $R_{K/\pi}$, signal 
  yield asymmetries and $N_{D^0K}$ is
  taken as a systematic uncertainty
  ($\pm 0.003$ for $A_{\CP+}$,
  $\pm 0.002$ for $A_{\CP-}$, $\pm 0.010$ for $R_{\CP+}$ and $\pm 0.007$ for $R_{CP-}$). 
  Possible \CP asymmetries up to $20\%$ in the peaking backgrounds are also taken
  into account ($\pm 0.04$ for $A_{\CP+}$).
  An estimate of the intrinsic detector charge
  bias due to acceptance, tracking, and particle identification
  efficiency has been obtained from the weighted average of the measured
  asymmetries in the processes $B^-{\ra}D^0[{\ra}K^-\pi^+]h^-$ and 
  $B^-{\ra}D^0_{\CP\pm}\pi^-$, where \CP\ violation is expected to be
  negligible. 
  This asymmetry estimate ($\pm 0.02$) has been added in quadrature to the total systematic
  uncertainties on the \CP\ asymmetries $A_{\CP\pm}$ (this is a correlated part of the systematics
  for $A_{\CP+}$ and $A_{\CP-}$).
  The accuracy in the equivalence between $R_{\pm}$ and
  $R_{\CP\pm}$ is evaluated to be $\pm 0.03$ for $R_{\CP+}$ and $\pm 0.02$ for $R_{\CP-}$
  (these uncertainties are correlated).

  \section{SUMMARY}
  \label{sec:Summary}
  In conclusion, we reconstruct \btodk\ decays with
  $D^0$ mesons decaying to non-\CP $K^\mp\pi^\pm$, \CP-even \KpKm and \pipi and
  \CP-odd $\KS \pi^0$ and $\KS \omega$ eigenstates. We have measured the \CP asymmetries
  $A_{\CP+} = 0.35\pm 0.09\stat\pm 0.05\syst$ and $A_{\CP-} = -0.19\pm
  0.12\stat\pm 0.02\syst$.
  Our result for $A_{\CP+}$ is $3.4\sigma$
  away from 0.  This constitutes the first evidence for direct {\CP}
  violation in $B^- \ra \Dz K^-$ decays.
  The double ratios of branching fractions are measured to be
  $R_{\CP+} = 1.07\pm 0.10\stat\pm 0.04\syst$ and $R_{\CP-} =
      0.81\pm 0.10\stat\pm 0.05\syst$.

  The corresponding values of $x_{\pm}$ and $r_B^2$ are extracted using equations 5 and 6,
  separately propagating correlated and uncorrelated errors on $A_{\CP\pm}$ and $R_{\CP\pm}$.
  We obtain
  $x_+ = -0.065 \pm 0.047\stat \pm 0.020\syst$,
  $x_- = 0.199 \pm 0.052\stat \pm 0.020\syst$, $r_B^2 = -0.060 \pm 0.070\stat \pm 0.039\syst$.

  The results obtained in this analysis are statistically in agreement with the previous measurements
  as demonstrated in Table~\ref{tab:comparison}.
  All results presented in this document are
  preliminary.

\begin{table}[!h]
    \caption{ \label{tab:comparison}
      Comparison of the preliminary results of this analysis 
      to the previous measurements by \babar~\cite{prl_babar_d0cpeven}
      and Belle~\cite{bib:Belle_BtoDcpK}. The decay mode $\Dz \ra \KS \phi$, used
      in the previous analyses, is not included in the present measurement. }
    \begin{center}
      \begin{tabular}{lccc}
        \hline
        \hline
        Parameter &\  Present analysis &\ \babar\ (2006)~\cite{prl_babar_d0cpeven} & Belle (2006)~\cite{bib:Belle_BtoDcpK} \\
        \hline
        $R_{\CP-}$ &\  $0.81  \pm 0.10 \pm 0.05$ &\ $ 0.86 \pm 0.10 \pm 0.05$ & $ 1.17 \pm 0.14 \pm 0.14$ \\
        $R_{\CP+}$ &\  $1.07  \pm 0.10 \pm 0.04$ &\ $ 0.90 \pm 0.12 \pm 0.04$ & $ 1.13 \pm 0.16 \pm 0.08$ \\
        $A_{\CP-}$ &\  $-0.19 \pm 0.12 \pm 0.02$ &\ $-0.06 \pm 0.13 \pm 0.04$ & $-0.12 \pm 0.14 \pm 0.05$ \\
        $A_{\CP+}$ &\  $0.35  \pm 0.09 \pm 0.05$ &\ $ 0.35 \pm 0.13 \pm 0.04$ & $ 0.06 \pm 0.14 \pm 0.05$ \\
        \hline
        \hline
      \end{tabular}
    \end{center}
  \end{table}

  \section{ACKNOWLEDGMENTS}
  \label{sec:Acknowledgments}
  We are grateful for the 
extraordinary contributions of our \pep2\ colleagues in
achieving the excellent luminosity and machine conditions
that have made this work possible.
The success of this project also relies critically on the 
expertise and dedication of the computing organizations that 
support \babar.
The collaborating institutions wish to thank 
SLAC for its support and the kind hospitality extended to them. 
This work is supported by the
US Department of Energy
and National Science Foundation, the
Natural Sciences and Engineering Research Council (Canada),
Institute of High Energy Physics (China), the
Commissariat \`a l'Energie Atomique and
Institut National de Physique Nucl\'eaire et de Physique des Particules
(France), the
Bundesministerium f\"ur Bildung und Forschung and
Deutsche Forschungsgemeinschaft
(Germany), the
Istituto Nazionale di Fisica Nucleare (Italy),
the Foundation for Fundamental Research on Matter (The Netherlands),
the Research Council of Norway, the
Ministry of Science and Technology of the Russian Federation, and the
Particle Physics and Astronomy Research Council (United Kingdom). 
Individuals have received support from 
CONACyT (Mexico),
the A. P. Sloan Foundation, 
the Research Corporation,
and the Alexander von Humboldt Foundation.

\begin{thebibliography}{99}
\bibitem{conj}
    Charge conjugation is implied here and throughout the text unless
    explicitly stated otherwise.
    \relax
\end{thebibliography}
\end{document}